# Handsets Malware Threats and Facing Techniques

Marwa M. A. Elfattah
Computer Science Dep,
Faculty of Computers and Information,
Helwan University, Cairo, Egypt

Aliaa A.A Youssif
Computer Science Dep,
Faculty of Computers and Information,
Helwan University, Cairo, Egypt

Ebada Sarhan Ahmed
Computer Science Dep,
Faculty of Computers and Information,
Helwan University, Cairo, Egypt

*Abstract*—Nowadays, mobile handsets combine the functionality of mobile phones and PDAs. Unfortunately, mobile handsets development process has been driven by market demand, focusing on new features and neglecting security. So, it is imperative to study the existing challenges that facing the mobile handsets threat containment process, and the different techniques and methodologies that used to face those challenges and contain the mobile handsets malwares. This paper also presents a new approach to group the different malware containment systems according to their typologies.

## I. INTRODUCTION

Recently, mobile handsets are becoming more intelligent and complex in functionality, much like PCs. Moreover, mobiles are more popular than PCs, and are being used more and more often to do business, access the Internet, access bank accounts, and pay for goods and services. This resulted in an increased number of criminals who wants to exploit these actions for illegal gains.

Today's malware is capable of doing many things, such as: stealing and transmitting the contact list and other data, locking the device completely, giving remote access to criminals, sending SMS and MMS messages etc. Mobile malware causes serious public concern as the population of mobile phones is much larger than the population of PCs [1-6].

The first proof-of-concept of mobile malware was "Cabir" [7], which was proposed in 2004 targeting Symbian OS. After that, mobile malware evolved rapidly during the first two years (2004 - 2006) of its existence. A wide range of malicious programs targeting mobile phones appeared, and these programs were very similar to malware which targeted computers: viruses, worms, and Trojans, the latter including spyware, backdoors, and adware. Now a days , the amount of malware for mobile devices have been duplicated more than one time. This shows that the growth rate demonstrated between 2004 and 2006 has been maintained.

In response to this increasing threat, this paper was developed as a survey to contain this critical phenomenon. The paper is organized to introduce the challenges of the mobile handsets malware containment and facing operation in section 2. Section 3 shows the different techniques which were proposed by researchers to face the mobile malware. Then, in section 4, a new approach to group the malware containment systems according to their typologies is introduced. To the best of authors' knowledge, this approach of grouping is never introduced before. Finally, in section 5, this work is concluded.

## II. MOBILE MALWARE FACING CHALLENGES

Although the great dangerous of mobile handset malware, and the importance of finding a solution to limit this danger, the task of facing mobile handset malware and limiting their harm has a lot of obstacles and is not easy to be faced. In this section most of those obstacles are concluded [1-6]:

- Some of mobile handset users treat mobile handset malware as a problem which has not happened yet, or believe that it's not an issue which really concerns them.

- A mobile handset has limited processing power and storage capacity, unlike resource-rich PCs, a detection framework should not consume too much of the device resources, including CPU, memory, and battery power, the overhead for executing the detection framework should be kept to a minimum.

- Most new malicious programs for mobile handset devices are hybrids, containing functionality from two or more different types of malware.

- When virus writers realized that there was no clear leading operating system for mobile devices, they also realized it wouldn't be possible to target the majority of mobile device users with a single attack. Because of this, they started focusing less on writing malware which targeted specific platforms, and more on creating programs capable of infecting several platforms.

- While a computer is primarily connected to the internet via IP networks, a mobile handset also connects to the cellular network through SMS/MMS services, as well as its Bluetooth interface that is frequently used to interact with other devices. These interfaces are quickly becoming the new infection vector for viruses, which makes the mobile handset susceptible to get infected even when it is disconnected from the internet.

- A mobile handset is highly mobile and always on, resulting in a greater degree of difficulty in quarantining the virus in a local region.

- To evade detection, malware writers are increasingly using polymorphic coding techniques. Polymorphism is a process through which malicious code modifies its appearance to evade detection without actually changing its underlying functionality. These techniques include everything from modifying the names of internal variables and subroutines, changing the order in which instructions appear in the body of malware, to encrypting most of the malware code, only leaving in the clear text the instructions necessary to decrypt the code [1]. In addition to changing the appearance of malware via polymorphism, new malware can further change their behavior, going through metamorphism; metamorphic code actually changes the functionality of malware, while hiding its payload using obfuscation and encryption [1]. When metamorphic techniques are used in conjunction with polymorphism, malware of this kind are much harder to detect, analyze, and filter.

### III. MOBILE MALWARE FACING TECHNIQUES

An effective detection frame work should be able to detect diverse types of malware and malware variants, keeping both false-negatives and false-positives below a certain acceptable threshold; also it should not consume the device resources.

There are set of approaches for preventing mobile handset from malware as shown in Fig 1. The simplest of them is to only trust and install digitally signed applications [4]. This ensures that the software has undergone a standard testing procedure as part of being signed. However, given the vast number of mobile applications available on the Internet, especially peer-to-peer sites, one cannot expect all applications to be signed with a certificate. An application that has been self-signed cannot be trusted to be free of malicious code. Moreover, even when an application is signed by a trusted CA, a malicious program can still enter the system via downloads (e.g., SMS/MMS messages with multimedia attachments), and it may exploit known vulnerabilities of an unsigned helper application.

Signature-based detection is another will known procedure for handling mobile handset malwares. It is relays on static file signature which make it vulnerable to simple obfuscation, polymorphism, and packing techniques. Also the needing of a huge database to store a signature for each malware makes this technique unsuitable for mobile devices which suffer from limited resource.

An alternative to signature-based methods is the behavioral-based detection, which has been emerged as a promising way of preventing the intrusion of spyware, viruses and worms.

#### A. Signature-Based Detection

Signature-based detection is one of the most known techniques for malware detection. To identify malwares, signature-based detection systems compare the contents of a file to a dictionary of malware signatures. There are well developed signature-based techniques for malware detection on the PC domain, but it requires considerable effort to adapt these techniques for mobile handsets. Also, signature-based detection techniques are unsuitable for mobiles because those techniques require a new signature for every single malware variant. However, mobile handsets have severe resource constraints in terms of memory and power.

Some of researchers attempted deal with these difficulties, and to adjust signature-bases detection algorithm to fit the mobile device. Deepak and Guoning [8, 9] suggested a system for detecting malware using malware signatures. This system automatically extracts a set of signatures from existing malware samples. It reduced the number of signatures by using a common signature for a malware and its variants.

Also, it minimized the total false alarm rate of malware detection by extracting signatures that are most uncommon within mobile network traffic. Deepak and Guoning outlined the considerations for malware detection on mobile devices and proposed a signature matching algorithm with low memory requirements and high scanning speed. They used hash table and sub-signature matching to scans the network traffic.

Although deepak solution consumes less than 50% of the memory used by Clam-AV while maintaining a fast scanning rate [8], signature-based detection methods still suffer from a lot of other weaknesses which make most of proposed signature-based solutions unsuitable for mobile handset [1, 4, 6]. Signatures are created using static information, thus being vulnerable to simple obfuscation, polymorphism, and packing techniques. Also, it is challenging to distribute virus signatures files to the mobile handsets in a timely manner. Even though, only one signature is required for a malware and its variants, the amount of those signatures is still more than the amount of malware behavior signatures, which are changed rarely.

Moreover, anti-virus solutions for mobile devices rely only on signature-based detection could not be considered as Conclusive solutions. Although the infected files are deleted by the anti-virus tool, the underlying vulnerability is not patched. As a result, a cleaned handset may get infected again by another instance of the same virus, requiring repeated cleanup.

#### B. Behavioral -Based Detection:

In behavioral-based detection techniques, the behavior of an application is monitored and compared against a set of malicious and/or normal behavior profiles. The malicious behavior profiles can be specified as global rules that are applied to all applications, as well as fine-grained application-specific rules [4-10].

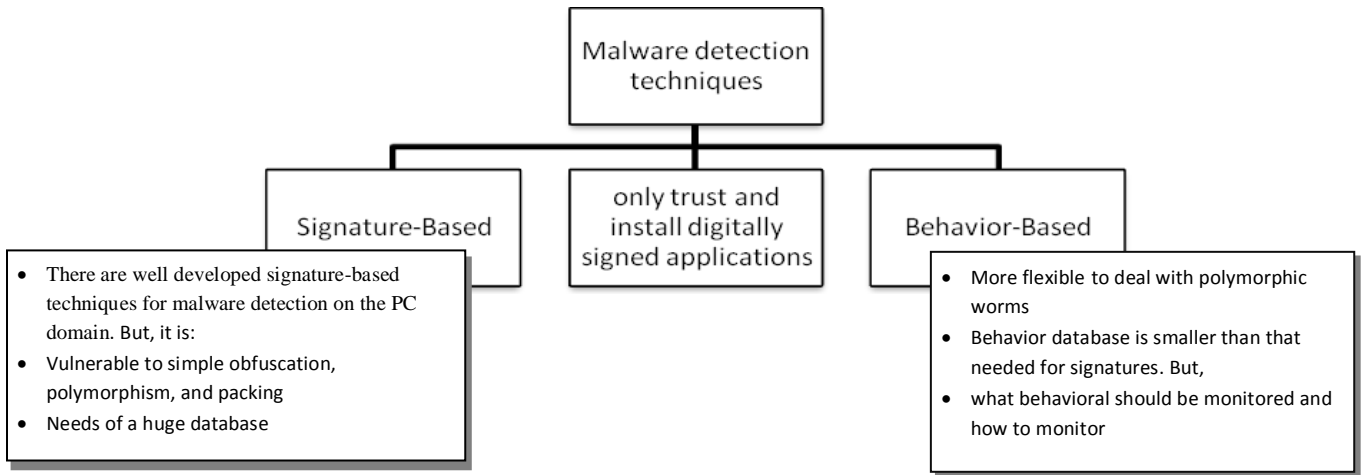

Figure 1: Mobile malware detection techniques

Behavioral detection is more flexible to deal with polymorphic worms or code obfuscation, because it assesses the effects of an application based on more than just specific payload signatures. Considering the fact that a new malware variant is usually created by adding new functionality to existing malware or modifying obsolete modules with fresh ones, this abstraction is effective for detecting previously-unknown malware variants that share a common behavior exhibited by previously-known malware. A typical database of behavior profiles and rules should be smaller than that needed for storing specific payload signatures of many different classes of malware. This makes behavioral detection methods particularly suitable for handsets.

One common problem with behavioral detection, however, is specification of what constitutes normal or malicious behavior that covers a wide range of applications [4], while keeping false positives (incorrect identification of a benign activity as malicious) and false-negatives (failure to identify malicious activities) low. Another one is the reconstruction method of potentially suspicious behavior from the applications, so that the observed signatures can be matched against a database of normal and malicious signatures. Heuristics on what behavioral should be monitored and how to monitor and collect behavioral vary. In the following, some of researchers' efforts to produce methodologies for malware detection and relevant behavioral measurements are concluded.

Code Analysis: There are several methods to analysis the executable code. For example, D. Venugopal et-el [11] proposed an algorithm to monitor each application and library (e.g., a dynamic link library (DLL)) that a process attempts to load. This information is then compared against lists of authorized and unauthorized applications and libraries.

They observed that most viruses in the mobile domain demonstrate common functionalities, such as, deleting system files and sending MMS messages, so mobile viruses are classified into different families or classes based on their functionalities. All virus variants in a family share a common malicious core behavior. Each malware needs to use certain dynamic link libraries (DLLs) to implement their functions. The DLL functions used by a virus give a good insight into the functionality of this virus. Therefore, the imported DLL functions were used as features for virus detection. They considered their method is computationally efficient since the DLL functions are easy to be extracted from the executable files [11].

Carsten Willems et-al developed CWSandbox [12], which is a malware analysis tool that employ dynamic malware analysis, API hooking and dynamic linked library (DLL) injection techniques to implement the necessary rootkit functionality to avoid detection by the malware. Using dynamic analysis techniques, they observed malware behavior and analyzes its properties by executing the malware in the sandbox. The analysis process can be done by taking an image of the complete system state before malware execution and comparing it to the complete system state after execution, or by monitoring the malwares actions during execution with the help of a specialized tool, such as a debugger.

To observe a given malware sample's control flow, it is important to access the application API functions. One possible way to achieve this is by hooking - intercepting a call to a function. When an application calls a function, it's rerouted to a different location where customized hook function resides. The hook then performs its own operations and transfers control back to the original API function or prevents its execution completely.

DLL code injection help in implementing of API hooking in a modular and reusable way. However, API hooking with inline code overwriting makes it necessary to patch the application after it has been loaded into memory. To be successful, the hook functions must be copied into the target application's address space so they can be called from within the target - this is the actual code injection - and bootstrap the API hooks in the target application address space using a specialized thread in the malware memory.

CWSandbox system outputs a behavior-based analysis that is; it executes the malware binary in a controlled environment so that it can observe all relevant function calls to the system API, and generates a high-level summarized report from the monitored API calls. To enable fast automated analysis, the CWSandbox is executed in a virtual environment so that the system can easily return to a clean state after completing the analysis process. But, this approach suffers from some drawbacks, such as, slower execution and device overloading.

Also, Liang Xie et-el [13] assumed that malware always launch attacks from the application software. From the application's point of view, malware attacks always cause anomalies in process states and state transitions. Such anomalies are reflected through malware function (API) calls, usages of system resources, and requests for system services. So, they adopt function call-trace techniques and human intelligence techniques in the context of cell phones to identify process misbehavior.

Also they noted that, each cell phone user has his own unique and private operational patterns (e.g., while operating keypad or touch-screen), which cannot be easily learned and simulated by malware. From these two aspects, and their behavior-based malware detection system (pBMDS) provides comprehensive protection against malware. pBMDS leverages a Hidden Markov Model (HMM) to learn process behaviors (states and state transitions) and additionally user operational patterns, such that it can effectively identify behavior difference between malware and human users for various cell phone applications.

File system Monitoring: A file system can be monitored through a number of aspects such as checking file integrity, file attributes, or file access attempts. Both file integrity and attribute checking can only be determined if a change has taken place [1], but file access attempts can be predetermined.

In checking for file access attempts, X. Zhang, et-al [14, 15] proposed a mandatory access control (MAC) system to strictly controls - according to some predefined sets of rules - the interactions between subjects (e.g., services or processes) and objects (e.g., files, sockets, etc.), which are differentiated based on the labels assigned to them. In the system an agent with a shim - a layer of code placed in between existing layers of code - can monitor all attempts to access critical files and stop suspicious attempts by comparing policies with the characteristics of the current attempt, such as which user / application attempts to access what file with a particular type of access (i.e., read, write, or execute).

The main advantage of this approach on mobile handsets is that, kernel-level mechanisms are intrinsically trusted, simply because that the kernel is a part of the trusted computing base. Also a MAC-based isolation is better than virtualization techniques due to the pure performance. Since mobile phones have limited computational capabilities and low power consumption requirements, virtualization becomes an impractical solution.

However, Although MAC mechanisms consume substantial computing power on PC platforms (due to vast number of subjects and objects), mobile handsets in contrast are still limited and cannot be compared to classical PC environments in this regard. This significantly simplifies the security policies and improves the potential performance of MAC mechanisms on mobile devices. But kernel-level solutions are too difficult to be implemented.

Power Consumption Monitoring: While most malicious code attacks on handhelds aim to damage software resources, intentional abuse of hardware resources (e.g., CPU, memory, battery power) has become an important, increasing threat. In particular, malware targeting the burning/depletion of battery power are extremely difficult to be detected and prevented, mainly because users are usually unable to recognize this type of anomaly on their handhelds and the battery can be deliberately and rapidly drained in a number of different ways.

H. Kim et-el [1] have designed a malware-detection framework, which is composed of a power monitor and a data analyzer. The former collects power samples and builds a power consumption history with the collected samples, and the latter generates a power signature from the power consumption history. The data analyzer then detects an anomaly by comparing the generated power signature with those in a database.

VirusMeter is another system that was developed by Liu L et-el [16] who illustrated that, by monitoring power consumption on a mobile device, VirusMeter catches misbehaviors that lead to abnormal power consumption. VirusMeter relies on a concise user-centric power model that characterizes power consumption of common user behaviors based on the number or the duration of the user actions, such as, the duration of Call, the number of SMS, and etc.

These works have shown that power anomaly is an effective indicator for suspicious activities on mobile phones. To identify the causes of these activities is still a challenge for power-based malware detection as the power consumption for normal behavior is yet to be accurately quantified. Another challenge is that existing mobile handsets is not able to provide sufficient precision for power consumption measurement without involving extra measuring devices like an oscilloscope [10].

Communicational Statistical Modeling: Statistical modeling for malware is usually used as a collaboration defense for preventing the malware spreading over the network. D. Venugopal, Hu. Guoning designed SmartSiren [2] which aims to detect worms exploiting SMS messaging and Bluetooth communication. This system keeps track of the communication activities on the device. In cases where abnormal activities have been locally identified, alerts are sent to both infected devices and a subset of the uninfected devices, which may be in contact with an infected device, based on the users' contact lists and mobility profiles.

IV. THE PROPOSED GROUPING APPROACH FOR MALWARE FACING METHODOLOGIES BASED ON TYPOLOGIES

To the best of authors' knowledge, there is not any study that groups the malware containment systems according to their typologies. But, it is widely found that, the done work on the field of mobile malware detection and prevention can be grouped into three complementary typologies, as shown in Fig 2.

### A. Device-Based Detection Typology:

Due to the danger of the malware attack which aim the mobile handsets, a lot of researchers have concentrate all of their attention on the device based solutions. They tried to face the malware attacks by proposing detection and prevention systems that are completely built on the device and they never affect or are affected by the infrastructure. For example, Aciicmez et-al [14] developed kernel level but general-purpose mandatory access control (MAC) mechanisms for main stream operating systems. Typically, a MAC system strictly controls the interactions between subjects (e.g., services or processes) and objects (e.g., files, sockets, etc.), which are differentiated based on the labels assigned to them.

Also, G Tuvell et-el[17] developed a system and method for detecting malware with in a device by modeling the behavior of malware and comparing a suspect executable with the model. The system and method extracts feature elements from malware-infected applications. Using malware-free and malware-infected applications as training data, the system and method heuristically trains the rules and creates a probability model for identifying malware. To detect malware, the system and method scans the suspect executable for feature sets and applies the results to the probability model to determine the probability that the suspect executable is malware-infected.

### B. Infrastructure-Based Detection Typology:

Another set of researchers noted that, each device affects and is affected by a not neglectable set of other devices, and the malware facing will be more proactive if it was in a collective manner instead of other individual solutions. Those researchers preferred to propose solutions that concern with the complete infrastructure. Their solutions are based on collecting information from the infrastructure components in organized manner. The collected information is used in the protection and the malware detection solutions.

For example, V. Karyotis et-el [18] have study the propagation of malware over a wireless ad hoc network. They proposed a probabilistic model that is able to model and capture the aggregated behavior of a large ad hoc network attacked by a malicious node, where legitimate network nodes are prone to propagate infections they receive to their neighbors. They used the Norton equivalent representation of the proposed network model that allowed them to acquire analytical results of the behavior of the system in its steady state. Depending on the acquired relations, they were able to identify the critical system parameters and the way they affect the operation of the network.

In order to analyze the influence of various system parameters on the network operation and identify which of them can be exploited by an attacker or by the network itself, they focused on the average number of infected nodes and the average throughput of the non-infected nodes, which could be indicative of the overall asymptotic system behavior independently of a specific network instant.

Furthermore, the Infection efficiency of an attack was obtained through simulation and used as a comprehensive attack evaluation metric in order to evaluate the impact of attackers on the network for a specific time period and scenario, indicating potential short-term variations and effects. Also, some insight regarding the behavior and evolution of the system when multiple attackers operate simultaneously and independently in the network is gained via modeling and simulation.

SmartSiren [2] is another example of infrastructure-based solution to securing mobile phones despite the limitations of post-infected detection. The goal of SmartSiren is to halt the potential virus outbreak by minimizing the number of mobile handsets that will be infected by a new released virus. The outbreak of viruses must affect many mobile handsets and cause noticeable changes in their behavior. Thus, early detection of viruses can be achieved by keeping track of the device activities even in a coarse granularity.

In this system, each mobile handset runs a light-weight agent, while a centralized proxy is used to assist the virus detection and alert processes. Each mobile handset agent keeps track of the communication activities on the device, and periodically reports a summary of these activities to the proxy. In cases where abnormal activities have been locally identified, a mobile handset may also submit a report immediately to the proxy. On the other hand, the proxy detects any single-device or system-wide viral behaviors. When a potential virus is detected, the proxy sends targeted alerts to both infected devices and a subset of the uninfected devices, which may be indirect contact with an infected device, based on the users' contact lists and mobility profiles.

For each user, based on the user-submitted communication log, the proxy would keep track of the average number of communications that each user initiates each day using a 7 days moving average window. The summation of the 7 days moving average captures the normal usage of each user and is considered as a threshold. In addition, each day the mobile handset user's agent will count the number of communication that the users have initiated. When the user's daily usage exceeds the threshold, the user would be moved from normal state into over-usage

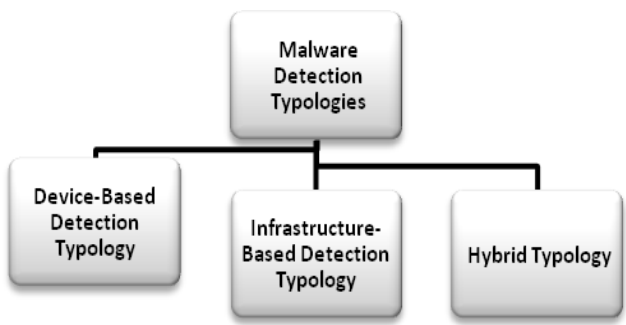

Figure 2: Malware Detection Typologies
The proposed grouping approach for malware facing methodologies based on typologies

46 | P a g e

state. The over-usage state does not guarantee that a particular handset is infected. The proxy also monitors how many users would exceed their threshold. When this daily count exceeds wildly from the average, it can suggest that an aggressive viral outbreak has occurred.

Also, Bose and Shin [3, 4, 6] discussed an agent-based malware modeling (AMM) framework that to investigate malware exploiting SMS/MMS and Bluetooth vulnerabilities on cellular handsets. In AMM, a mobile network was modeled as a collection of autonomous decision-making entities called agents. The agents represent networked devices within the network such as PDAs, mobile phones, service centers and gateways. In case of agents representing mobile devices, the connectivity changes as users roam about the physical space of the network. The behaviors of the agents are specified by a set of services running on them.

Thus, there are two types of topologies in Bose and Shin simulation environment. The physical connectivity, whereas the logical connectivity is determined by the messages exchanged among the agents. An agent may participate in multiple logical topologies corresponding to different services like email, IM, SMS, etc. They also group the agents in a hierarchical manner. For example, agents representing cellular base stations can keep track of mobile devices in their respective cells. Accordingly, these agents are able to collect information aggregated over the individual devices in their respective cells. This capability of higher-level agents to aggregate observations collected from lower-level agents reflects real-life processing of information within a mobile network. The information processed at these different levels can also be used to activate different response mechanisms against a spreading malware.

C. *Hybrid Typology:*

As it is illustrated previously, mobile handsets have only limited resources in terms of computation, storage, battery power. Also mobile handset users always annoy from any slowing down on the mobile performance, so, the user should never be disturbed with the existence of the detection systems. Those limitations restrict the device base solutions and harden their rule. Although Infrastructure-based solutions are computationally more expensive than device-based solutions, they offload most of the processing burden from the resource-constrained mobile handsets, thus minimizing the performance penalty on the mobile handsets.

Also, it is important, for accurate virus detection and prompt alerts, the mobile handsets must collaborate with each other. Infrastructure-based solutions simplify this collaboration among the mobile handsets by using a centralized agent that serves such collaboration. Of course, the centralized agent marked itself as a performance bottleneck and a single point of failure in the system. To improve the scalability and resiliency, one can extend the architecture with multiple agents.

Some of researchers have proposed a hybrid approach for malware detection and prevention. On this approach, part of the detection system is placed in the device, and the other part is placed on the infrastructure. For example, A. Schmidt et-al [19] introduces an approach of how to monitor mobile handsets in order to extract values that can be used for remote anomaly detection. Therefore, it has to be learned what is the normal behavior of a user and device in order to be able to distinguish between normal and abnormal, possibly malicious actions. The extracted features are sent as vector to a remote system, taking the responsibility for extended security measures away from the probably unaware user. These vectors can be used for methods from the field of artificial intelligence in order to detect abnormal behavior.

Another framework was proposed by H. Kim et-al [1], which was composed of a power monitor and a data analyzer. The former on the device collects power samples and builds a power consumption history with the collected samples, and the latter on the remote server generates builds a power signature from the power consumption history. The data analyzer then detects an anomaly by comparing the generated power signature with those in a database.

Also, Michael Becher [20] has said that, a promising approach is an automatic dynamic analysis, where system calls are logged and afterwards analyzed for malicious behavior. Because of mobile handset limitations, this cannot be done efficiently on the mobile handset. They designed a Mobile Sandbox to analysis the collected samples in a mobile dynamic malware analysis system. It executes the sample in an environment (the sandbox), where it can watch the steps of the investigated sample. An important requirement to ensure the integrity of the analysis is logging to a remote place rather than saving the log on the device only. Mobile Sandbox implements this communication of the device to the host system with a TCP connection over ActiveSync.

V. CONCLUSION

Due to their flexible communication and computation capabilities, and their resource constraints, mobile handsets are glued victim to malwares. A mobile handset can be attacked from the Internet since mobile are Internet endpoints, or it can be Infected from compromised PC during data synchronization; also it can have a peer mobile attack or infection through SMS/MMS and Bluetooth.

Although the difficulties of building a malware detection systems, some of researchers have concerned with this field of research. Set of researcher have done work to adjust the existed PC's signature-based detection systems to be suitable for mobiles. But, signature-based solutions have a lot of weakness, so another set of researchers preferred behavioral based solutions due to their flexibility to deal with polymorphic worms, and to the small amount of data needed to be stored. However, on behavioral based solutions, it is important to specify what constitutes normal or malicious behavior that covers a wide range of applications. This paper presented and analyzed some of researchers' effort on that aspect

On other hand, according to detection system typologies, this paper has grouped the detection systems into three complementary typologies, which are device-based detection typology, infrastructure-based detection typology. Device-based detection typology faces the resource limitation constrains, and has no way that facilitate the communication and the collaboration between detection systems. However infrastructure-based detection typology is computationally more expensive and the centralized agent marked itself as a performance bottleneck and a single point of failure in the system. In hybrid topologies, a part of the detection system is placed in the device, and the other part is placed on the infrastructure.

REFERENCES

[1] H. Kim, J. Smith, G. Shin, "Detecting energy-greedy anomalies and mobile malware variants", The International Conference on Mobile Systems, Applications, and Communications (MobiSys), ACM/USENIX, pp. 239–25, 2008.

[2] J. Cheng, S. Wong, H. Yang, Lu. Songwu, "Smartsiren: virus detection and alert for smartphones", The International Conference on Mobile Systems, Applications, and Communications , MobiSys, pp. 258-271, ACM, Jun. 2007.

[3] A. Bose, G. Shin., "On mobile viruses exploiting messaging and bluetooth services", International Conference on Security and Privacy in Communication Networks ,SecureComm, IEEE, PP. 1-10, Aug. 2006.

[4] A. Bose, "Propagation, detection and containment of mobile malware", PhD thesis , the university of Michigan, 2008.

[5] G. Chuanxiong, J. Wang, Z. Wenwu, "Smart-phone attacks and defenses", Third Workshop on Hot Topics in Networks, HotNets III, San Diego, CA, 2004.

[6] A. Bose, X. Hu Kang G. Shin, T. Park, "Behavioral detection of malware on mobile handsets", International Conference on Mobile Systems, Applications, and Communications , MobiSys08, pp. 225-238, June, 2008.

[7] http://www.viruslist.com/en/analysis.

[8] D. Venugopal, Hu. Guoning, "Efficient signature based malware detection on mobile devices", Mobile Information Systems journal, Vol. 4, pp. 33-49, 2008.

[9] Hu. Guoning, V. Deepak, "A Malware signature extraction and detection method applied to mobile networks", Performance, Computing, and Communications Conference, IPCCC 2007, IEEE Internationa, pp. 19 – 26, 2007.

[10] Q. Yan, R. H. Deng, Yingjiu Li, and Tieyan Li, "On the potential of limitation-oriented malware detection and prevention techniques on mobile phones", International Journal of Security and Its Applications, Vol. 4, No. 1, pp.21-30, January, 2010.

[11] D. Venugopal, Hu. Guoning, N. Roman, "Intelligent virus detection on mobile devices", Fourth International Conference on Privacy, Security and Trust, ACM PST, pp. 1-4, 2006.

[12] C. Willems, T. Holz, F. Freiling, "Toward automated dynamic malware analysis using cwsandbox", IEEE Security & Privacy, pp. 32-39, March 2007.

[13] L. Xie, X. Zhang, J. Seifert, S. Zhu, "PBMDS: A behavior-based malware detection system for cellphone Devices", 3rd ACM conference on wireless network security,WiSec10, ACM, pp. 37-48, March, 2010.

[14] X. Zhang, Aciicmez, O. Latifi, A. Seifert, S. Jose, "A rusted mobile phone prototype", Consumer Communications and Networking Conference, CCNC 2008. 5th IEEE, pp. 1208—1209, 2008.

[15] X. Zhang, L. Xie, A. Chaugule, T. Jaeger, S. Zhu, "Designing system-level defenses against cellphone malware", 28th IEEE International Symposium on Reliable Distributed Systems, pp.83-90, 2009.

[16] Liu L., Yan, G., Zhang, X., Chen, S., "VirusMeter: preventing your cellphone from spies", Proceedings of the 12th International Symposium on Recent Advances in Intrusion Detection, pp. 244-264, 2009.

[17] G. Tuvell, D. Venugopal, Hu. Guoning, "Malware modeling detection system and method for mobile platforms", freepatentsonline.com, 2007.

[18] V. Karyotis, A. Kakalis, and S. Papavassiliou, "Malware-propagative mobile ad hoc networks: asymptotic behavior analysis", Journal Of Computer Science And Technology, vol 23, pp.389-399, May 2008.

[19] A. Schmidt, F. Peters, F. Lamour, "Monitoring smartphones for anomaly detection", Mobile Networks and Applications Volume 14, Mobilware08, ACM, Number 1, pp. 92-106, 2008.

[20] M. Becher, F. C. Freiling, "Towards Dynamic Malware Analysis to Increase Mobile Device Security", SICHERHEIT, 2008

[21] Mobile Malware Evolution: An Overview, Part 3. http://www.securelist.com/en/analysis?pubid=204792080